\documentclass[namedreferences]{SolarPhysics}
\usepackage[optionalrh]{spr-sola-addons} 
\usepackage{graphicx}
\usepackage{color}
\usepackage{url}             

\newcommand{\apj}{    {\it Astrophys. J.}}
\newcommand{\apjl}{   {\it Astrophys. J. Lett.}}

\newcommand{\jgr}{    {\it J. Geophys. Res.}}

\newcommand{\solphys}{{\it Solar Phys.}}

\newcommand{\ssr}{    {\it Space Sci. Rev.}}

\begin{document}
\begin{article}
\begin{opening}

\title{A Simple Way to Estimate the Soft X-ray Class of Far-Side
Solar Flares Observed with STEREO/EUVI }

\author{I.M.~\surname{Chertok}$^{1}$\sep
        A.V.~\surname{Belov}$^{1}$\sep
        V.V.~\surname{Grechnev}$^{2}$
        }

\runningauthor{Chertok et al.}
 \runningtitle{Soft X-ray Class of Far-Side Flares}

\institute{${}^{1}$Pushkov Institute of Terrestrial Magnetism,
            Ionosphere and Radio Wave Propagation (IZMIRAN), Troitsk, Moscow, 142190 Russia email: \url{ichertok@izmiran.ru}; \url{abelov@izmiran.ru} \\
 ${}^{2}$Institute of Solar-Terrestrial Physics SB RAS,
            Lermontov St.\ 126A, Irkutsk 664033, Russia email: \url{grechnev@iszf.irk.ru}
                       }

\date{Received ; accepted }

\begin{abstract}
Around the peaks of substantial flares, bright artifact nearly
horizontal saturation streaks (B-streaks) corresponding to the
brightest parts of the flare sources appear in the STEREO/EUVI
195~\AA\ images. We show that the length of such B-streaks can be
used for the solution of an actual problem of evaluating the soft
X-ray flux and class of far-side flares registered with double
STEREO spacecraft but invisible from Earth. For this purpose from
data on about 350 flares observed from January 2007 to July 2014
(mainly exceeding the GOES M1.0 level) both with GOES and STEREO,
an empirical relation is established correlating the GOES
1--8~\AA\ peak flux and the B-streak length. This allowed us for
the same years to estimate the soft X-ray classes for
approximately 65 strong far-side flares observed by STEREO. The
results of this simple and prompt method are consistent with the
estimations of Nitta {\it et al.} (Solar Phys., 288, 241, 2013)
based on the calculations of the EUVI full-disk digital number
output. In addition, we studied some features of the B-streaks in
impulsive and long-duration flares and demonstrated that B-streaks
in several consecutive EUVI images can be used to reconstruct a
probable time history of strong far-side flares.

\end{abstract}
\keywords{STEREO; Extreme Ultraviolet; Solar Flares; Saturation Streaks;
GOES; Soft X-Rays}

\end{opening}

\section{Introduction}
  \label{S-Introduction}

In October 2006, the double \textit{Solar Terrestrial Relations
Observatory} (STEREO; \opencite{Kaiser2008}) was launched, and at
the end of January 2007 the two spacecraft separated and entered
into heliocentric orbits near 1~AU in opposite directions. The
Ahead (A) probe leads the Earth, while the Behind (B) probe lags
behind the Earth, drifting about $22^{\circ}$ per year from the
Sun--Earth line. In February 2011, the two STEREO spacecraft were
already in the quadrature with the Earth, providing the first ever
complete $360^{\circ}$ view of the Sun. At the end of July 2014,
the two spacecraft were located on the opposite sides of the
Earth's orbit; STEREO-A was ahead of the Earth by $164^{\circ}$,
and STEREO-B was $162^{\circ}$ behind the Earth. Each STEREO
spacecraft is equipped with an almost identical set of extreme
ultraviolet, optical, radio, and in situ instruments. In
particular, the \textit{Sun Earth Connection Coronal and
Heliospheric Investigation suit} (SECCHI; \opencite{Howard2008})
includes the \textit{Extreme Ultraviolet Imager} (EUVI;
\opencite{Wuelser2004}) providing solar images in four channels of
171, 195, 284, and 304~\AA.

With an increase of the longitudinal separation between the two
STEREO spacecraft and each of them with the Sun--Earth line, the
number of flares registered with the A and/or B probes but
invisible from Earth has been growing. Among the problems in
studies of the backside flares is classifying their importance,
which would allow one to compare them with flares recorded by
near-Earth satellites. STEREO observations and information about
the importance of far-side flares are significant for
investigations of such solar activity phenomena as flares
themselves, eruptions, large-scale coronal waves, coronal mass
ejections (CMEs), solar energetic particles (SEPs). These
observational studies are promising to achieve further progress in
understanding and modeling these phenomena as well as space
weather forecasting (e.g., \opencite{Hudson2011};
\opencite{Lugaz2012}; \opencite{Webb2012}; \citeauthor{Nitta2013a}
\citeyear{Nitta2013a, Nitta2013b}; \opencite{Aschwanden2014};
\opencite{Li2014}; \opencite{Richardson2014};
\opencite{Kwon2015}). Currently, in addition to the classification
of flares based on the intensity and emission area in the
H$\alpha$ line, the soft X-ray flare (SXR) classification is
generally accepted and widely used. The C, M, and X SXR classes of
flares are determined according to the peak fluxes measured by the
\textit{Geostationary Operational Environmental Satellite} (GOES;
\opencite{Garcia1994}) 1--8~\AA\ detectors in the ranges of
(1--10) $\times 10^{-6}$, $10^{-5}$, and $10^{-4}$~W~m$^{-2}$,
respectively.

\inlinecite{Nitta2013a} addressed the problem of classification of
the STEREO backside flares calculating the EUVI 195~\AA\ full-disk
emission fluxes as a total output of the charge-coupled device (CCD)
camera in units of digital data number (DN) per second. Using data
from June 2010 to September 2012, the authors selected the flares
that were recorded by GOES and simultaneously were observed with one
or both STEREO spacecraft and found an empirical relation between
the observed GOES 1--8~\AA\ peak fluxes and the calculated EUVI
195~\AA\ full-disk DN output. Using this relation, one can estimate
the GOES peak X-ray flux of sufficiently intense backside flares
observed only by STEREO. \inlinecite{Nitta2013a} presented a list of
16 such major far-side flares detected with this procedure.

\begin{figure} 
  \centerline{\includegraphics[width=1.0\textwidth]
   {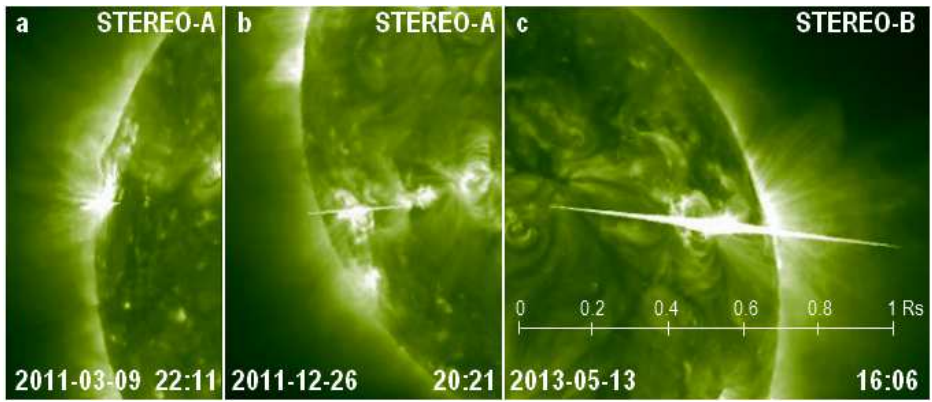}
  }
  \caption{The STEREO/EUVI 195~\AA\ B-streaks typical of C, M, and X class flares.
  The spatial scale shown in panel (c) is the same for all of the three images.
  The 2011-03-09, 2011-12-26, and 2013-05-13 flares are labeled 77, 157, and 295
  in Figure~\ref{F-scatter_plot} (see also Table~\ref{T-GOES_STEREO}).}
  \label{F-different_flares}
  \end{figure}

In the present paper, we propose a somewhat simpler technique for
estimations of the X-ray class of STEREO flares. The technique is
based on the measurements of the length of a bright artifact
saturation streak, which appears in the EUVI 195~\AA\ images
nearly along its East--West axis around the peaks of sufficiently
strong flares (Figure~\ref{F-different_flares}). Such a streak is
a consequence of the so-called blooming, \textit{i.e.}, saturation
of CCD cells, corresponding to the brightest part of a flare
source, and spilling of excessive electrons from these cells along
CCD columns (see \opencite{Wuelser2004}). Similar overexposure
effects occur also in the \textit{Extreme ultraviolet Imaging
Telescope} (EIT; \opencite{Delab1995}) images gathered with SOHO
(\textit{e.g.}, \opencite{Andrews2001}) and in the
\textit{Atmospheric Imaging Assembly}  (AIA; \opencite{Lemen2012})
images obtained with SDO \cite{Schwartz2014}.

Henceforth we will refer to the blooming streaks as B-streaks. The
B-streaks affect the images of flare cores around the peak time.
These sources and moments are most interesting physically. The
B-streaks are considered usually as a serious interference for the
image processing and studies. Our analysis shows however that
B-streaks contain useful information about the flare importance and
the brightest EUV sources in general. This idea is based on the fact
that the larger emission flux produces the stronger blooming in CCD
cells, and therefore a longer B-streak is formed (see
Figure~\ref{F-different_flares}). In the next
section~\ref{S-Approach}, we describe our approach and data selected
for the analysis. Section~\ref{S-Relations} is devoted to strong
flares observed with both GOES and one or two STEREO spacecraft.
These concurrent observations allowed us to obtain an empirical
relation between the relative lengths of B-streaks and the SXR
fluxes (\textit{i.e.}, the GOES classes) of the STEREO flares. In
Section~\ref{S-Strong_Flares}, we present illustrations and a list
of major backside flares (mainly above the M3.0 class), registered
by STEREO spacecraft in 2007--2014. We also compare the SXR classes
of far-side flares evaluated by \inlinecite{Nitta2013a}, using the
calculated DN fluxes in the EUVI 195~\AA\ channel, and those
estimated with our technique. The paper ends in
Section~\ref{S-Summary} with a summary and concluding remarks.

\section{Approach and Data}
  \label{S-Approach}

In order to judge to what extent B-streaks are suitable for an
assessment of the SXR class of backside flares, it is necessary to
analyze the relationship between these parameters for coincident
flares observed simultaneously with GOES and one or both STEREO
spacecraft. Proceeding from the STEREO trajectories, it is clear
that initially the number of such coincident flares was large, but
then gradually decreased. After February 2011, when the two STEREO
spacecraft were in the near-quadrature configuration relative to
the Sun--Earth line, only those flares were observed both with
GOES and one of the STEREO probes that were sufficiently removed
from the central meridian visible from Earth. In 2013 and 2014,
only near-the-limb flares, from the Earth view, could be observed
with GOES and one of the STEREO spacecraft; the west- and
east-limb GOES flares were observed with A and B stations,
respectively.

We have analyzed the NOAA GOES flare list up to July 2014,
\textit{i.e.}, during the ascent and maximum phases of Solar Cycle
24,
(\url{ftp://ftp.ngdc.noaa.gov/STP/space-weather/solar-data/solar-features/solar-flares/x-rays/goes/})
and selected for further analysis practically all coincident flares
above the M1.0 class. The exceptions were a small number of flares
overlapping in time, but occurring in different active regions;
flares with obvious partial limb occultation for either STEREO or
GOES; and some M1--M2 flares, in which a relatively short B-steak
was not discernible against extended pre-existing structures. In
addition to strong flares, a certain amount of C-class flares with a
visible B-streak was randomly chosen to be included in our analysis.
We considered these events to reveal more clearly the tendency of
the increasing length of the EUV B-streak with an increasing SXR
flare importance.

For the considerations of B-streaks we used EUVI images in the
195~\AA\ channel, which encompasses the Fe~{\sc xii} line with a
peak temperature of $\approx 1.5$~MK, because these images were
produced mostly every 5 min (sometimes the cadence was 2.5 or 10
min). More importantly, this channel has some sensitivity to the
emission of high-temperature plasma due to the secondary peak around
15~MK in the Fe~{\sc xxiv} line at 192~\AA\ that makes the 195~\AA\
channel the closest analog of the SXR GOES monitors (see
\opencite{Nitta2013a}). As the main quantitative parameter
characterizing B-streaks, we took its maximum length, $L$, measured
in fractions of the solar radius, $R_\mathrm{S}$, in the same EUVI
image, \textit{i.e.}, the $L/R_\mathrm{S}$ ratio. As will be shown
later, the length of a B-streak can considerably exceed
$1R_\mathrm{S}$ in major flares.

We endeavored to develop a method, which would be as simple as
possible. For this reason, we did not analyze the source FITS files,
and used instead the EUVI 195~\AA\ images in JPEG and movies in
MPEG, which are available at
\url{http://stereo-ssc.nascom.nasa.gov/browse/}. They are already
processed with the standard SolarSoft routine secchi\_prep.pro, and
rotated so that solar north is up. In spite of their non-linear
brightness scale, most B-streaks are clearly discernible in the $512
\times 512$ movies.  We also use the $2048 \times 2048$ images to
measure the B-streak length, especially for weak flares. The
observed B-streak length is expected to depend on the exposure time,
$\tau_\mathrm{exp}$. For each image, information on
$\tau_\mathrm{exp}$ can be found at
\url{http://sharpp.nrl.navy.mil/cgi-bin/swdbi/secchi_flight/img_short/form},
using the full-page output. It turned out that more often the EUVI
telescopes operated with $\tau_\mathrm{exp} = 8$~s. The 16~s
exposure time was almost regularly used in early observations until
January 2010. Later on, $\tau_\mathrm{exp} = 16$~s had only the
images recorded every even hour at the 16th minute, \textit{i.e.},
at 00:16, 02:16, 04:16, and so on (\textit{all times hereafter refer
to UT}). We corrected all the considered images to
$\tau_\mathrm{exp} = 8$~s. Namely, if the longest B-streak occurred
in an image with $\tau_\mathrm{exp} = 16$~s, we halved its length
and compared it with B-streaks in adjacent images. We then excluded
flares (mainly of C-class) in which after such a correction the
maximum relative length of B-streaks, $L/R_\mathrm{S}$, was less
than 0.1. The reason is that short B-streaks in such events could
become visible against the background of other flare structures
just because $\tau_\mathrm{exp} = 16$~s.

\begin{table}
 \caption{Extraction from the list of flares registered both with
GOES and STEREO (see section~\ref{S-Approach}). Asterisks mark the
events in which the correction for an exposure time of 8~s was
applied.}
 \label{T-GOES_STEREO}
 \begin{tabular}{rcccccrccc}
 \hline

&\multicolumn{6}{c}{GOES SXR} & \multicolumn{2}{c}{STEREO} \\

\multicolumn{1}{c}{No.} & \multicolumn{1}{c}{Date} & \multicolumn{1}{c}{Time} & \multicolumn{1}{c}{Class} &
 \multicolumn{1}{c}{Flux} & \multicolumn{1}{c}{Location} & \multicolumn{1}{c}{AR} &  \multicolumn{1}{c}{A/B} & \multicolumn{1}{c}{Time} & \multicolumn{1}{c}{$L/R_\mathrm{S}$} \\

 &  & \multicolumn{1}{c}{UT} & \multicolumn{1}{c}{} & \multicolumn{1}{c}{$F_\mathrm{G}$} &  &
 \multicolumn{1}{c}{} &  & \multicolumn{1}{c}{UT} & \\

\hline

\multicolumn{1}{c}{(1)} & \multicolumn{1}{c}{(2)} & \multicolumn{1}{c}{(3)} & \multicolumn{1}{c}{(4)} &
\multicolumn{1}{c}{(5)} & \multicolumn{1}{c}{(6)} & \multicolumn{1}{c}{(7)} & \multicolumn{1}{c}{(8)} &
\multicolumn{1}{c}{(9)} & \multicolumn{1}{c}{(10)} \\

 \hline

17 & 2010-02-06 & 18:59 & M2.9 & 29 & N21\,E17 & 11045 & A & 19:03 & 0.34 \\

27 & 2010-02-12 & 11:26 & M8.3 & 83 & N26\,E11 & 11046 & B & 11:26 & 0.54 \\

77$^*$ & 2011-03-09 & 22:12 & C9.4 & 9.4 & N08\,W04 & 11166 & A & 22:11 & 0.11 \\

78 & 2011-03-09 & 23:23 & X1.5 & 150 & N08\,W09 & 11166 & A & 23:26 & 0.27 \\

99 & 2011-07-30  & 02:09 & M9.3 & 93 & N16\,E19 & 11261 & B & 02:11 & 0.60 \\

107 & 2011-08-09 & 08:05 & X6.9 & 690 & N17\,W69 & 11263 & A & 08:06 & 1.59 \\

112 & 2011-09-06 & 22:18 & X2.1 & 210 & N14\,W18  & 11283 & A & 22:21 & 0.61 \\

113 & 2011-09-07 & 22:38 & X1.8 & 180 & N14\,W28 & 11283 & A & 22:41 & 0.61 \\

121 & 2011-09-22 & 11:01 & X1.4 & 140 & N13\,E78 & 11302 & B & 10:56 & 0.19 \\

123 & 2011-09-24 & 09:40 & X1.9 & 190 & N12\,E60 & 11302 & B & 09:41 & 0.77 \\

142 & 2011-11-03 & 20:27 & X1.9 & 190 & N22\,E63 & 11339 & B & 20:26 & 0.53 \\

157 & 2011-12-26 & 20:30 & M2.3 & 23 & S21\,W42 & 11387 & A & 20:21 & 0.26 \\

171 & 2012-01-27 & 18:37 & X1.7 & 170 & N27\,W71 & 11402 & A & 18:51 & 0.40 \\

177$^*$ & 2012-03-05 & 04:09 & X1.1 & 110 & N16\,E54 & 11429 & B & 04:16 & 0.25 \\

189$^*$ & 2012-03-07 & 00:24 & X5.4 & 540 & N17\,E27 & 11429 & B & 00:26 & 0.66 \\

233 & 2012-07-06 & 23:08 & X1.1 & 110 & S18\,W52 & 11515 & A & 23:11 & 0.42 \\

275 & 2012-10-23 & 03:17 & X1.8 & 180 & S13\,E52 & 11598 & B & 03:21 & 0.47 \\

294$^*$ & 2013-05-13 & 02:17 & X1.7 & 170 & N12\,E78 & 11748 & B & 02:16 & 0.42 \\

295$^*$ & 2013-05-13 & 16:01 & X2.8 & 280 & N11\,E85 & 11748 & B & 16:06 & 0.95 \\

296 & 2013-05-14 & 01:11 & X3.2 & 320 & N13\,E81 & 11748 & B & 01:11 & 0.77 \\

297 & 2013-05-15 & 01:48 & X1.2 & 120 & N12\,E64 & 11748 & B & 01:46 & 0.37 \\

307 & 2013-10-25 & 08:01 & X1.7 & 170 & S04\,E72 & 11882 & B & 08:03 & 0.38 \\

308 & 2013-10-25 & 15:03 & X2.1 & 210 & S03\,E68 & 11882 & B & 15:01 & 0.64 \\

315$^*$ & 2013-10-28 & 02:03 & X1.0 & 100 & N04\,W66 & 11875 & A & 02:06 & 0.24 \\

319 & 2013-10-29 & 21:54 & X2.3 & 230 & N08\,W86 & 11875 & A & 21:56 & 0.58 \\

321 & 2013-11-19 & 10:26 & X1.0 & 100 & S13\,W79 & 11893 & A & 10:26 & 0.38 \\

337 & 2014-02-25 & 00:49 & X4.9 & 490 & S12\,E82 & 11990 & B & 00:51 & 0.56 \\

346 & 2014-04-25 & 00:27 & X1.3 & 130 & S15\,W90 & 12046 & A & 00:26 & 0.59 \\

347 & 2014-06-10 & 11:42 & X2.2 & 220 & S15\,E80 & 12087 & B & 11:41 & 0.27 \\

348 & 2014-06-10 & 12:52 & X1.5 & 150 & S17\,E82 & 12087 & B & 12:56 & 0.23 \\

 \hline
 \end{tabular}
 \end{table}

The list of events selected for our analysis contains about 275
flares, whose GOES importance was M1 or higher, and 75 C-class
flares. The complete list of flares observed with both GOES and
STEREO/EUVI is accessible at
\url{http://www.izmiran.ru/~ichertok/STEREO/}.
Table~\ref{T-GOES_STEREO} presents some flares extracted from this
list and contains, in particular, those events that will be mentioned
below, as well as a number of major flares. Column 1 presents the
serial number of a flare in the complete list. Columns 2--7 contain
information on GOES SXR observations of a flare in 1--8~\AA,
including date; peak time; COES class; corresponding SXR flux,
$F_\mathrm{G}$, in units of $10^{-6}$~W~m$^{-2}$; coordinates; and
the NOAA number of an active region. Columns 8--10 specify STEREO A
or B spacecraft; the observation time of the B-streak (rounded to
1~min); and its maximum relative length, $L/R_\mathrm{S}$.

\section{Relation between the GOES Flux and EUVI B-streak}
 \label{S-Relations}

In the preceding section, was mentioned how to correct for the
EUVI exposure time. Some other factors can affect the relations
between the lengths of the EUVI B-streaks and the GOES SXR fluxes.
These are:

-- Different temperature responses of the SXR GOES detectors and
the EUVI 195~\AA\ channel, as mentioned in
section~\ref{S-Approach}.

-- Possible time difference between the flare peak in SXR and EUV
emissions.

-- Limited imaging rate of the EUVI (section~\ref{S-Approach}) and
possible related omissions of flare peaks.

-- Significant differences between impulsive (compact) flares and
long-duration events (LDEs, which are usually associated with big
CMEs; see below).

-- A difference (though small) between the distances of the
STEREO-A and B spacecraft from the Sun.

-- Particularities of the B-streak formation in the blooming
process.

-- We measure the longest B-streak only, although several
B-streaks occur in some flares (mainly LDEs; see below).

\begin{figure} 
  \centerline{\includegraphics[width=0.9\textwidth]
   {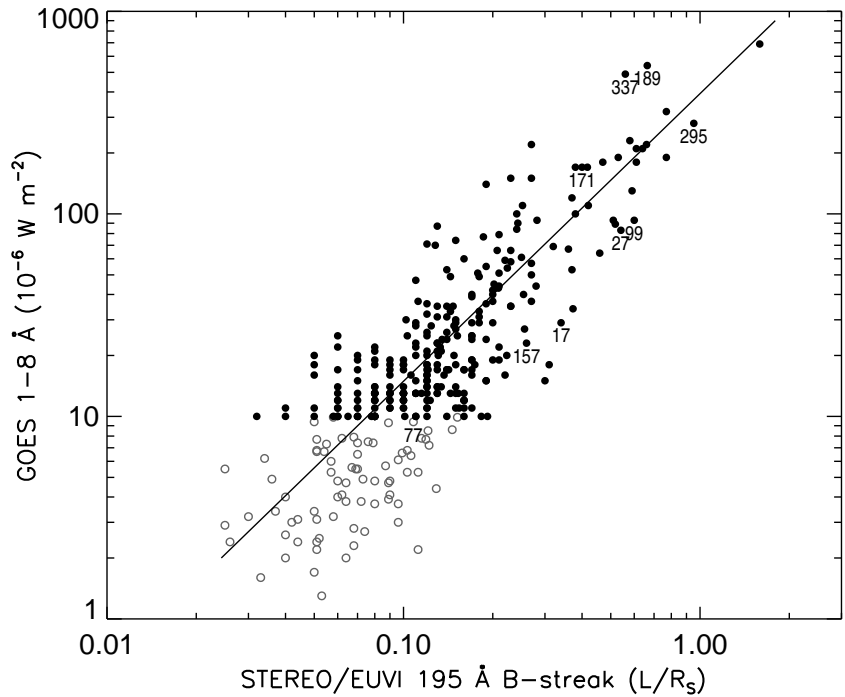}
  }
  \caption{Scatter plot of the relative lengths of the STEREO/EUVI
B-streaks versus the GOES 1--8~\AA\ fluxes. The gray open circles
denote C-class flares, and the black filled circles denote
$\geq$~M1-class flares. The line corresponds to the regression
equation (\ref{E-regression}). The numbers specify the flares
shown in Figures \ref{F-different_flares}, \ref{F-imp_flares}, and
\ref{F-LDE_flares}.}
  \label{F-scatter_plot}
  \end{figure}

We disregard these factors for simplicity of the method.
Nevertheless, as Figure~\ref{F-scatter_plot} shows, a clear
relationship does indeed exist between the maximum relative length
of the B-streak, $L/R_\mathrm{S}$, and the peak GOES 1--8~\AA\
flux, $F_\mathrm{G}$. On average, when $L/R_\mathrm{S}$ increases
from 0.03 to 1.5, $F_\mathrm{G}$ rises from 3 to 600 (hereafter
$F_\mathrm{G}$ is expressed in the unit of $10^{-6}$~W~m$^{-2}$)
that corresponds to the SXR flare class from C3 to X6. The C-class
flares confirm the trend which M- and X-class flares show. We
remind that only a limited number of C flares is presented here
which were selected randomly and had a conspicuous B-streak. The
visibility of short B-streaks is obviously hampered, because
pre-flare and flaring structures have some longitudinal extent,
too. Therefore, the points of C-class flares in
Figure~\ref{F-scatter_plot} are biased to relatively longer
B-streaks due to their visual selection and the limited number of
events. Further we consider only the most powerful flares above
the M1.0 level. Because we do not control any of the variables in
the $F_\mathrm{G} - L/R_\mathrm{S}$ relationship, the geometric
mean regression method was chosen to obtain an empirical
dependence between these quantities. Using this method, we
obtained a power-law regression equation to fit the relation
between the length of the B-streak and the SXR flux,
\begin{eqnarray}
F_\mathrm{G} = 392 \times (L/R_\mathrm{S})^{1.42}
 \label{E-regression}
\end{eqnarray}
(the line in Figure~\ref{F-scatter_plot}). The correlation
coefficient between $L/R_\mathrm{S}$ and $F_\mathrm{G}$ is
$r\approx 0.81$. For the majority of the $\gsim $\,M1.0 flares,
deviations of the SXR flux from the regression line down or up do
not exceed a factor of 2. Equation (\ref{E-regression}) can be
used for estimations of the SXR fluxes and GOES classes of
far-side flares observed by STEREO but invisible from Earth. We
will return to this issue in section~\ref{S-Strong_Flares}.

\begin{figure} 
  \centerline{\includegraphics[width=1.0\textwidth]
   {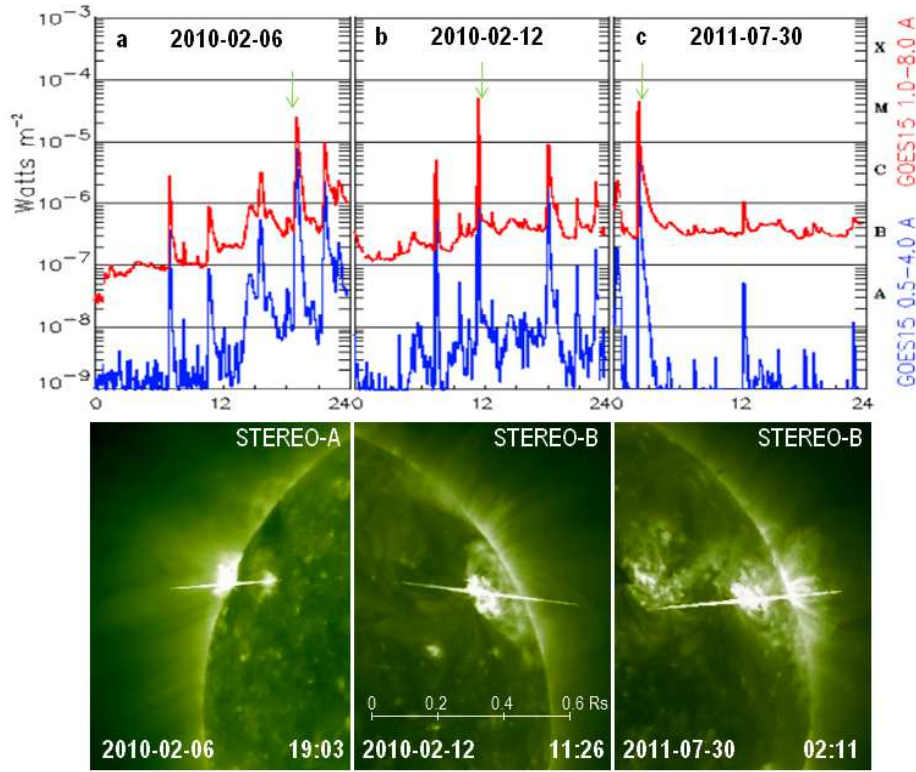}
  }
  \caption{Impulsive GOES M-class flares (upper row) and
their thin, single, long B-streaks (bottom). The 2010-02-06,
2010-02-12, and 2011-07-30 events are labeled 17, 27, and 99 in
Figure~\ref{F-scatter_plot} (see Table~\ref{T-GOES_STEREO}).}
  \label{F-imp_flares}
  \end{figure}

Our analysis shows that many impulsive and LDE flares are
characterized by different B-streaks. As Figure~\ref{F-imp_flares}
demonstrates, in the standard 3-day GOES plots, impulsive flares
look like spikes almost without any distinct decay phase. In such
flares, the whole duration and decay times measured when the flux
level decays to a point halfway between the maximum flux and the
pre-flare background level, are 10--20 min and 3--10 min,
respectively. Such a B-streak is visible usually only in 2--3 EUVI
frames of the 5-min cadence.  Figure~\ref{F-imp_flares}
illustrates that impulsive flares produce a single, thin and
relatively long B-streak (see Figure~\ref{F-scatter_plot} and
Table~\ref{T-GOES_STEREO}). This is due to the fact that almost
the entire flux from such flares is emitted by a single compact
core corresponding to a few pixels. Accordingly, such events
reside under the regression line in Figure~\ref{F-scatter_plot}.

\begin{figure} 
  \centerline{\includegraphics[width=1.0\textwidth]
   {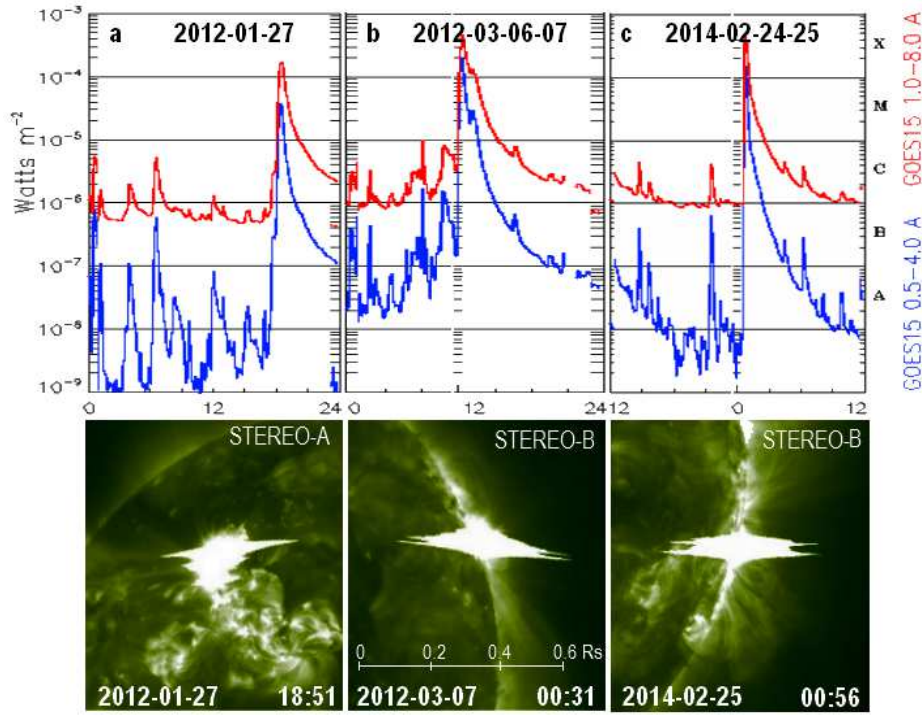}
  }
  \caption{Long-duration GOES X-class flares (upper row) and their thick,
two or multi-element, relatively short B-streaks (bottom). The
2012-01-27, 2012-03-07, and 2014-02-25 events are labeled 171,
189, and 337 in Figure~\ref{F-scatter_plot} (see
Table~\ref{T-GOES_STEREO}).}
  \label{F-LDE_flares}
  \end{figure}

In contrast, LDE flares produce relatively short, but thick
B-streaks with two or more blooming elements
(Figure~\ref{F-LDE_flares}). Apparently, a number of bright source
kernels corresponding to these blooming elements contribute to their
peak fluxes. In its most developed form, such a multi-element
blooming structure does not always temporally coincide to the
appearance of the longest B-streak. In the event shown in
Figure~\ref{F-LDE_flares}a, several short blooming elements almost
merged into one thick B-streak, while in the flares presented in
Figures \ref{F-LDE_flares}b and \ref{F-LDE_flares}c, the thick part
of the B-streak and diffuse core are contained between two thin
blooming elements. The LDE flares have noticeably longer durations
of 35--75 min and decay times of 15--45 min; their multi-element
B-streaks are visible in larger numbers of EUVI images (up to 10).
Let us remind that in such events we take into account only one, the
longest B-streak. Feasible summation of the lengths of all visible
B-streaks is not always unambiguous, and would complicate the method
proposed. The fact that we consider only one streak can be the
reason why the mentioned LDE events are located above the regression
line in Figure~\ref{F-scatter_plot}.

\begin{figure} 
  \centerline{\includegraphics[width=0.6\textwidth]
   {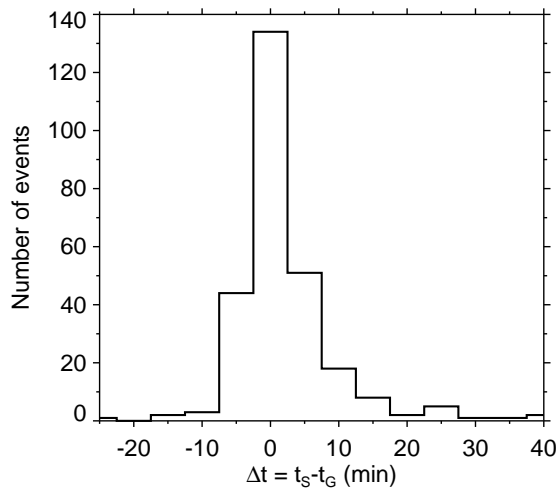}
  }
  \caption{The distribution of the time difference between
the observations of the longest STEREO/EUVI B-streak,
$t_\mathrm{S}$, and the peak of the SXR flux measured by GOES,
$t_\mathrm{G}$, for coincident $\geq$ M1.0 flares.}
  \label{Delay}
  \end{figure}

Based on the fact that the longest B-streaks indicate the
brightest flare sources, it is interesting to consider the time
difference, $\Delta{t}$, between the observation of such a
B-streak in a STEREO/EUVI image, $t_\mathrm{S}$, and a flare peak
in the soft X-ray GOES data, $t_\mathrm{G}$. The corresponding
histogram for $\geq$~M1.0 flares is presented in
Figure~\ref{Delay}. The time difference does not exceed $\pm
5$~min in the majority of events, being mainly due to the limited
imaging rate of EUVI. Practically in all flares with
$L/R_\mathrm{S}\geq 0.35$, the longest B-streak was observed near
the GOES SXR peak. Obviously, $|\Delta{t}|$ should be small in
short-duration flares. Pre-flare heating of brightened erupting
filaments or their static extensions produce earlier B-streaks
($\Delta{t} < 0$).

On the other hand, there is a noticeable percentage of LDEs, in
which the longest EUV B-streaks were observed with a considerable
delay after the SXR peak. Delayed B-streaks with $\Delta{t}\geq
10$~min occurred a few times more often than advancing B-streaks
with $\Delta{t}\leq -5$~min. Such multiple, relatively short
B-streaks in LDEs are produced by many loops of large long-lived
arcades. The situation in LDEs is more complex than in impulsive
events. Some delayed B-streaks can probably be caused by local
brightenings due to the overlap of slowly moving flare loops. Some
other delayed B-streaks can be due to subsidiary flare episodes on
the background of the long-lasting post-eruption energy release
late in such events, almost all of which are associated with big
CMEs.

\section{Powerful Far-Side Flares}
 \label{S-Strong_Flares}

In this section, we present powerful far-side flares which were
not visible from Earth but were found due to B-streaks through a
visual inspection of daily STEREO/EUVI movies and images of
2007--2014. Almost no such far-side flares were found in
2007--2009 for two reasons. Firstly, major flares occurred rarely
during this period of low activity. Secondly, two parts of the
solar surface invisible from Earth but observed with STEREO
spacecraft were still small. The longitudinal extent of these
parts gradually increased, and early in 2011 the whole far-side of
the Sun became accessible to STEREO observations. In 2014, both
STEREO probes were located almost exactly behind the Sun, and
almost all far-side flares were observed by the two spacecraft
simultaneously.

  \begin{table}
 \caption{A list of major far-side flares detected from B-streaks
in STEREO/EUVI 195~\AA\ images (see Section~4). Asterisks in column
6 mark the events in which the correction for an exposure time of
8~s was applied.}
 \label{T-strong_flares}
 \begin{tabular}{rcccccrcc}
 \hline

& & & & \multicolumn{5}{c}{Estimated SXR}  \\
\multicolumn{1}{c}{Date} & \multicolumn{1}{c}{Time} & \multicolumn{1}{c}{A/B} & \multicolumn{1}{c}{Location} &
\multicolumn{1}{c}{AR} & \multicolumn{1}{c}{$L/R_\mathrm{S}$} & \multicolumn{1}{c}{Flux} & \multicolumn{1}{c}{Class} & \multicolumn{1}{c}{Nitta \textit{et al.}}\\
& \multicolumn{1}{c}{UT} &  &  &\multicolumn{1}{c}{} & &
\multicolumn{1}{c}{$F_\mathrm{G}$} & & \multicolumn{1}{c}{class range}  \\

\hline
\multicolumn{1}{c}{(1)} & \multicolumn{1}{c}{(2)} & \multicolumn{1}{c}{(3)} & \multicolumn{1}{c}{(4)} &
\multicolumn{1}{c}{(5)} & \multicolumn{1}{c}{(6)} & \multicolumn{1}{c}{(7)} & \multicolumn{1}{c}{(8)} &
\multicolumn{1}{c}{(9)} \\
\hline

2010-01-17 & 03:56 & B & S25\,E128 & 11041 & 0.54 & 163 & X1.6 & M3.4--M9.6\\

2010-01-19 & 13:41 & B & S25\,E95 & 11041 & 0.33 & 81.2 & M8.1 & --\\

2010-01-19 & 20:36 & B & S25\,E92 & 11041 & 0.37 & 95.6 & M9.6 & --\\

2010-07-29 & 08:31 & A & N29\,W136 & -- & 0.23 & 48.6 & M4.9 & --\\

2010-08-31 & 20:56 & A & S22\,W146 & 11100 & 0.49 & 142.4 & X1.4 & M8.4--X2.5\\

2010-09-01 & 21:51 & A & S22\,W162 & 11100 & 0.42 & 114.4 & X1.1 & M5.4--X1.6\\

2010-11-03 & 12:16 & B & S19\,E98 & 11121 & 0.26$^*$ & 57.8 & M5.8 & --\\

2011-03-21 & 02:16 & A & N17\,W129 & 11169 & 0.05$^*$ & 5.6 & C5.6 & M1.3--X1.3\\

2011-06-03 & 21:06 & A & N15\,W143 & 11222 & 0.24 & 51.7 & M5.2 & --\\

2011-06-04 & 07:06 & A & N15\,W144 & 11222 & 0.32 & 77.7 & M7.8 & M5.2--X1.6\\

2011-06-04 & 21:51 & A & N17\,W148 & 11222 & 1.20 & 508 & X5.1 & X4.0--X12\\

2011-08-31 & 20:06 & A & N23\,W119 & 11278 & 0.30 & 70.9 & M7.1 & --\\

2011-09-03 & 02:46 & A & N22\,W148 & 11278 & 0.31 & 74.3 & M7.4 & --\\

2011-10-18 & 23:46 & A & N15W\,130 & -- & 0.33 & 81.2 & M8.1 & --\\

2011-10-19 & 06:21 & A & N14\,W134 & -- & 0.27$^*$ & 61.1 & M6.1 & --\\

2011-10-20 & 03:21 & A & N18\,W101 & 11318 & 0.20 & 39.9 & M4.0 & \\

2011-10-23 & 23:16 & A & N19\,W151 & 11318 & 0.32 & 77.7 & M7.8 & M5.3--X1.6\\

2011-10-26 & 22:57 & B & N18\,E167 & 11339 & 0.42 & 114 & X1.1 & --\\

2011-11-03 & 22:41 & B & N09\,E154 & -- & 0.09 & 12.8 & M1.3 & M4.7--X1.4\\

2012-03-26 & 22:56 & B & N18\,E123 & 11451 & 0.49 & 142.4 & X1.4 & M8.2--X2.5\\

2012-04-26 & 15:51 & A & S23\,W123 & 11462 & 0.26 & 57.8 & M5.8 & --\\

2012-04-27 & 16:06 & A & S22\,W136 & 11462 & 0.32 & 77.7 & M7.8 & --\\

2012-04-29 & 12:46 & B & N12\,E163 & 11467 & 0.77 & 270 & X2.7 & M8.3--X2.5\\

2012-06-12 & 05:06 & A & S15\,W126 & 11495 & 0.22 & 45.7 & M4.6 & --\\

2012-07-02 & 08:36 & B & S16\,E133 & 11520 & 0.26 & 57.8 & M5.8 & --\\

2012-07-15 & 15:51 & A & S20\,W173 & 11515 & 0.22 & 45.7 & M4.6 & --\\

2012-07-23 & 02:36 & A & S15\,W133 & 11520 & 0.18 & 34.3 & M3.4 & M8.2--X2.5\\

2012-08-19 & 18:16 & A & S22\,W178 & 11538 & 0.24$^*$ & 51.7 & M5.2 & --\\

2012-08-21 & 20:16 & B & S22\,E158 & 11538 & 0.28$^*$ & 64.3 & M6.4 & M6.8--X2.0\\

2012-09-11 & 07:56 & B & S22\,E178 & 11576 & 0.90 & 338 & X3.4 & X1.3--X3.9\\

2012-09-12 & 22:56 & B & S21\,E157 & 11576 & 0.30 & 70.9 & M7.1 & --\\

2012-09-14 & 20:52 & B & N09\,E140 & 11575 & 0.27 & 61.1 & M6.1 & --\\

2012-09-15 & 11:26 & B & N09\,E131 & 11575 & 0.24 & 51.7 & M5.2 & --\\

2012-09-19 & 11:16 & A & S16\,E169 & 11582 & 0.51 & 151 & X1.5 & M9.1--X2.7\\

2012-09-20 & 15:01 & B & S15\,E156 & 11582 & 2.38 & 1343 & X13 & X5.8--X18\\

2012-09-22 & 03:06 & B & S16\,E135 & 11582 & 0.38 & 99.2 & M9.9 & M8.4--X2.5\\

2012-09-27 & 10:36 & A & S24\,W154 & 11574 & 0.30 & 70.9 & M7.1 & --\\

2012-10-08 & 13:56 & B & S29\,E107 & 11590 & 0.50 & 146 & X1.5 & --\\

2012-10-19 & 17:31 & A & S21\,W156 & 11582 & 0.21 & 42.7 & M4.3 & --\\

 \hline
 \end{tabular}
 \end{table}

\setcounter{table}{1}

 \begin{table}
 \caption{\textit{Continued}}
 \begin{tabular}{rcccccrcc}
 \hline

& & & & \multicolumn{5}{c}{Estimated SXR}  \\
\multicolumn{1}{c}{Date} & \multicolumn{1}{c}{Time} &
\multicolumn{1}{c}{A/B} & \multicolumn{1}{c}{Location} &
\multicolumn{1}{c}{AR} & \multicolumn{1}{c}{$L/R_\mathrm{S}$} & \multicolumn{1}{c}{Flux} & \multicolumn{1}{c}{Class} & \multicolumn{1}{c}{Nitta \textit{et al.}}\\
& \multicolumn{1}{c}{UT} &  &  &\multicolumn{1}{c}{} & &
\multicolumn{1}{c}{$F_\mathrm{G}$} & &
 \multicolumn{1}{c}{class range}  \\

 \hline

\multicolumn{1}{c}{(1)} & \multicolumn{1}{c}{(2)} & \multicolumn{1}{c}{(3)} & \multicolumn{1}{c}{(4)} &
\multicolumn{1}{c}{(5)} & \multicolumn{1}{c}{(6)} & \multicolumn{1}{c}{(7)} & \multicolumn{1}{c}{(8)} &
\multicolumn{1}{c}{(9)} \\
\hline

2013-04-24 & 21:46 & A & N09\,W169 & 11719 & 0.23 & 48.6 & M4.9 & --\\

2013-05-01 & 02:31 & B & N15\,E115 & 11739 & 0.32 & 77.7 & M7.8 & --\\

2013-10-05 & 06:56 & B & S23\,E125 & 11865 & 0.23 & 48.6 & M4.9 & --\\

2013-10-31 & 20:25 & A & N12\,W139 & 11880 & 0.51 & 151 & X1.5 & --\\

2013-11-02 & 04:26 & A & N03\,W139 & 11875 & 0.83 & 301 & X3.0 & --\\

2013-11-04 & 05:21 & A & N01\,W165 & 11875 & 0.21 & 42.7 & M4.3 & --\\

2013-11-21 & 00:56 & A & S23\,W123 & 11901 & 0.42 & 114 & X1.1 & --\\

2013-11-21 & 16:26 & A & S22\,W132 & 11901 & 0.25 & 54.7 & M5.5 & --\\

2014-01-06 & 07:56 & A & S15\,W113 & 11936 & 0.67 & 222 & X2.2 & --\\

2014-01-26 & 08:37 & B & S16\,E107 & 11967 & 0.42 & 114 & X1.1 & --\\

2014-01-31 & 15:07 & B & S14\,E155 & 11974 & 0.30 & 70.9 & M7.1 & --\\

2014-02-10 & 18:16 & A & N13\,W161 & -- & 0.34$^*$ & 84.7 & M8.5 & --\\

2014-02-11 & 13:26 & A & N09\,W112 & 11968 & 0.46 & 130 & X1.3 & --\\

2014-02-14 & 01:36 & A & N08\,W137 & 11968 & 0.28 & 64.3 & M6.4 & --\\

2014-02-14 & 08:26 & A & S13\,W143 & 11967 & 0.39 & 103 & X1.0 & --\\

2014-02-20 & 03:16 & A & S17\,E142 & 11990 & 0.25 & 54.7 & M5.5 & --\\

2014-03-04 & 18:26 & A & N13\,W171 & -- & 0.74 & 256 & X2.6 & --\\

2014-03-05 & 13:26 & A & N13\,W179 & -- & 1.43 & 651 & X6.5 & --\\

2014-03-06 & 12:21 & A & N12\,E166 & 12007 & 0.34$^*$ & 84.7 & M8.5 & --\\

2014-03-09 & 02:51 & A & N17\,W159 & 11986 & 0.23 & 48.6 & M4.9 & --\\

2014-05-01 & 09:31 & A & N03\,E148 & 12056 & 0.78 & 275 & X2.8 & --\\

2014-05-09 & 02:31 & A & S11\,W119 & 12051 & 0.35 & 88.3 & M8.8 & --\\

2014-06-16 & 02:16 & A & S13\,W103 & 12080 & 0.20$^*$ & 39.9 & M4.0 & --\\

2014-06-17 & 09:01 & A & S13\,W123 & 12080 & 0.28 & 64.3 & M6.4 & --\\

2014-06-30 & 17:01 & B & S21\,E125 & -- & 0.26 & 57.8 & M5.8 & --\\

2014-09-01 & 11:01 & B & N14\,E129 & 12158 & 0.39 & 103 & X1.0 & --\\

2014-09-01 & 22:21 & B & S14\,E121 & 12157 & 0.22 & 45.7 & M4.6 & --\\

 \hline
 \end{tabular}
 \end{table}

The SXR fluxes (GOES classes) of such backside flares were
estimated from the relative lengths $L/R_\mathrm{S}$ of the
longest B-streaks using equation (\ref{E-regression}). We
restricted ourselves to flares with $L/R_\mathrm{S} \geq 0.2$,
which corresponds to the GOES class of $\gsim$~M4.0. The results
are presented in Table~\ref{T-strong_flares}, where for each of
the detected events the following information is listed: the date
(column 1); time of the longest B-streak (2); the STEREO A or B
probe, in whose images the streak was measured (3); approximate
coordinates of the flare (4); the NOAA number of active region (5)
assigned before its disappearance behind the west limb (mainly for
A-probe) or after its appearance at the east limb (mainly for
B-probe); the value of the $L/R_\mathrm{S}$ parameter (6); the
estimated peak SXR flux in units of $10^{-6}$~W~m$^{-2}$ (7), and
the flare GOES class (8). In accordance with the small statistical
errors of the parameters found in evaluating equation
(\ref{E-regression}), the 95\% confidence interval of the
estimated flare class is rather narrow for short B-streaks and
broadens gradually with an increase of $L/R_\mathrm{S}$. For
example, the confidence interval is M2.6--M4.7 for an M3.5 flare
and X10--X18 for an X13 flare.

Column 9 of Table~\ref{T-strong_flares} presents probable ranges of
the GOES classes for major far-side flares of 2010--2012 estimated
by \inlinecite{Nitta2013a} from the calculated total DN output. We
included all of the 16 backside flares listed by the authors,
although three of these flares had $L/R_\mathrm{S} < 0.2$. In two of
these events (2011-03-21, 02:11 and 2011-11-03, 22:41), the longest
B-streaks were even shorter, $L/R_\mathrm{S} < 0.1$, but their
blooming consisted of many elements, that is typical for LDE flares
associated with large CMEs. The third flare of 2012-07-23, 02:36
with $L/R_\mathrm{S} \approx 0.18$ is particularly noteworthy and
will be discussed below. As Table~\ref{T-strong_flares} shows, our
analysis of EUVI B-streaks revealed 63 far-side flares above the
M4.0 level, including 22 X-class flares, in addition to 94 and 28
flares of the same classes registered by GOES during the ascending
and maximum phases of the current solar cycle (until September
2014). Table~\ref{T-strong_flares} demonstrates that our estimations
of the GOES classes for 13 other major backside flares listed by
\inlinecite{Nitta2013a} are close to their results as well. In the
majority of cases, our estimated GOES importance falls within the
range specified by these authors.

\begin{figure} 
  \centerline{\includegraphics[width=1.0\textwidth]
   {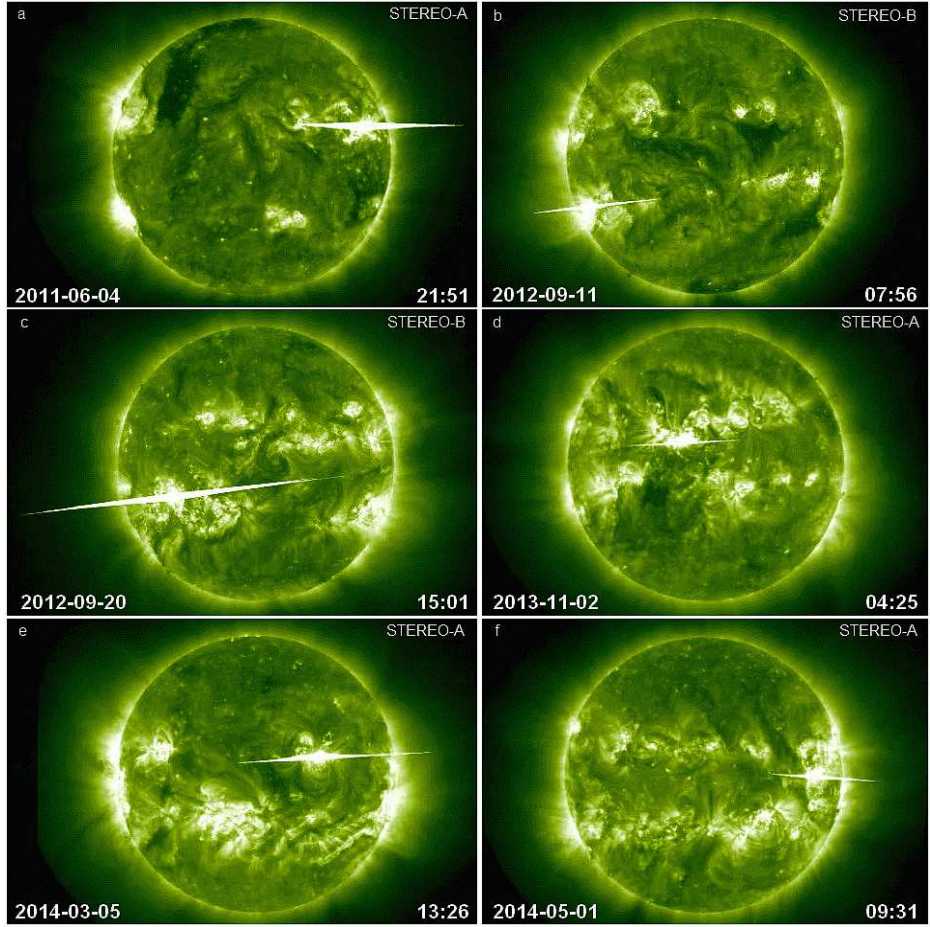}
  }
  \caption{Full-Sun STEREO/EUVI 195~\AA\ images of six strongest
far-side flares detected from B-streaks (see
Table~\ref{T-strong_flares}).}
  \label{F-major_flares}
  \end{figure}

\begin{figure} 
  \centerline{\includegraphics[width=1.0\textwidth]
   {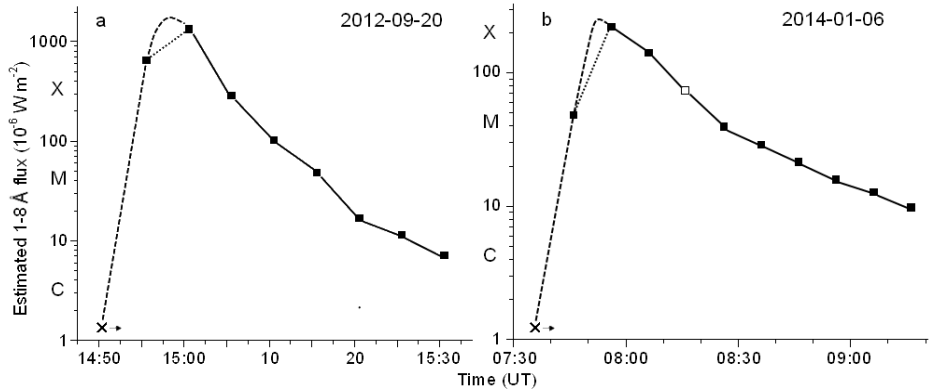}  }
  \caption{Probable time history of the 1--8~\AA\ flux of two famous
far-side flares estimated from STEREO/EUVI B-streaks (see the text
and Table~\ref{T-strong_flares}). The crosses with the right
arrows correspond to the images in which B-streaks are not yet
visible. The dashed lines represent a possible ascending part of
the time profiles. The open square in panel (b) refers to an image
corrected for an exposure time of 8~s.}
  \label{F-time_profiles}
  \end{figure}

Figure~\ref{F-major_flares} illustrates B-streaks in six of the
strongest backside flares observed by STEREO/EUVI. Shown in
Figure~\ref{F-major_flares}c is the 2012-09-20 flare with the
longest B-streak exceeding the solar diameter; in this case
$L/R_\mathrm{S} \approx 2.38$ that corresponds to the estimated
SXR class of X13. This event turned out to be not only the most
powerful far-side flare but the strongest flare of the Solar Cycle
24 until the present time. This conclusion and evaluation of the
flare class are consistent with the results of
\inlinecite{Nitta2013a} obtained through the calculations of the
full-disk EUVI total DN output. With an example of this event, we
can demonstrate that B-streaks allow not only to estimate the SXR
importance of large LDE flares but also to reconstruct their
probable time history. For this purpose, it is sufficient to have
visible B-streaks in several images, to measure their largest
relative length, and to calculate the SXR flux using equation
(\ref{E-regression}). As Figure~\ref{F-time_profiles}a shows, the
B-streaks in the 2012-09-20 event are clearly visible, at least,
in eight frames of the 5-min cadence. At the growth phase, a long
B-streak appears very sharply. The shape of the time profile
indicates that the maximum flux occurred between two frames of
14:56 and 15:01. This suggests that the SXR class of this flare
was still higher than the longest B-streak implies and seems to be
about X20.

Two other far-side events induced interest as sources of
outstanding space weather disturbances. The 2012-07-23 flare
mentioned above is widely debated and considered as an utmost
extreme eruptive event (\textit{e.g.}, \opencite{Ngwira2013};
\opencite{Liu2014}; \opencite{Temmer2015}). The corresponding
interplanetary CME (ICME) arrived at STEREO-A, which was located
favorably, in a very short transit time of 19--21~hr and brought
to 1~AU a record strong interplanetary magnetic field of 109~nT.
If directed toward Earth, this ICME might cause a geomagnetic
storm comparable to the famous Carrington storm of 1859.
Meanwhile, the corresponding flare was fairly moderate;
\inlinecite{Nitta2013a} estimated its soft X-class to be in the
M8.2--X2.5 range, and our estimation from the observed B-streak
length leads to even lower values of about M5.4. The researchers
of this event suggested that the extreme characteristics of the
event were caused by a high initial CME speed (up to
3000~km~s$^{-1}$), week ICME drag deceleration in the solar wind
determined by preceding solar eruptions, and by an enhancement of
the magnetic field due to an interaction between two CMEs, which
closely followed each other. Currently we are working on an
alternative interpretation suggesting that the extremely high
speed of CME/ICME and its strong interplanetary magnetic field
were due to a large eruptive magnetic flux and a weak ICME
expansion in propagation from the Sun to 1 AU, in accordance with
patterns described by \inlinecite{Chertok2013} and
\inlinecite{Grechnev2014}. The magnetic flux was estimated from
the SDO/HMI magnetogram on 2012-07-12, when the parent active
region was located near the center of the visible disk.

The 2014-01-06, 07:56 flare occurred at a heliolongitude of W113 and
produced a 2.5\% ground-level enhancement (GLE) of cosmic rays, only
the second one in the unusual Solar Cycle 24.
\inlinecite{Thakur2014} analyzed this proton event and kinematics of
an accompanying CME and came to a conclusion that it was consistent
with a particle acceleration by a CME-driven shock. On the other
hand, a sufficiently long ($L/R_\mathrm{S} \approx 0.67$)
multi-element B-streak existed in STEREO-A images for nearly one
hour during this flare. A probable time history presented in
Figure~\ref{F-time_profiles}b demonstrates that the flare was a
powerful LDE of $\approx$~X2 class. According to
\inlinecite{Belov2007}, an X2 class flare can well be associated
with a source of SEP with proton fluxes of J($>10$ MeV) $\approx
120$ pfu, J($>100$ MeV) $\approx 4.5$ pfu, and GLE of $\approx 2 \%$
that are close to the observations. These parameters estimated from
B-streaks are typical of flares related to small GLEs.

\section{Summary and Concluding Remarks}
 \label{S-Summary}

We have demonstrated how a spurious instrumental effect, which
strongly interferes solar flare imaging, can be used to obtain a
useful information. Here we were dealing with the so-called
blooming, arising from a significant saturation of the STEREO/EUVI
195~\AA\ images in CCD cells corresponding to the cores of
sufficiently strong flares. The saturated CCD cells lose their
ability to accommodate any additional charge, causing them to spill
over adjacent cells. As a result, one or several bright nearly
horizontal streaks (B-streaks) are formed. We have shown that in
spite of many unaccounted factors, the maximum relative length of
the EUV B-streak, $L/R_\mathrm{S}$, correlates with the peak flare
SXR flux, $F_\mathrm{G}$, measured by GOES. We have analyzed about
350 flares of 2007--2014 which were observed simultaneously by GOES
and one of STEREO spacecraft, and found an empirical relation
between $L/R_\mathrm{S}$ and $F_\mathrm{G}$. This allowed us to
propose a simple and prompt method for estimating the SXR class of
far-side flares observed with STEREO but invisible for GOES.

The method consists of a direct measurement of the length of the
longest B-streak in units of the solar radius in the routine EUVI
195~\AA\ images or movies, accessible in near real time at
\url{http://stereo-ssc.nascom.nasa.gov/browse/}, and the use of
equation (\ref{E-regression}). It is necessary only to make sure
that all the measurements were converted to the exposure time of
8~s. Prior to this, \inlinecite{Nitta2013a} proposed a somewhat
more laborious method for estimations of the SXR flux and class of
far-side flares based on calculations of the EUVI full-disk total
digital number output. It should be noted that if there is a
temporal overlap between the flares in different active regions,
then our method makes it possible to measure their B-streaks and
to estimate their classes independent of each other.

Applying our method to the STEREO observations during the ascending
and maximum phases of Solar Cycle 24 allowed us to find about 65
major backside flares with the EUVI B-streak lengths of
$L/R_\mathrm{S} \geq 0.2$ and to estimate their SXR fluxes and
classes. For the majority of 16 backside flares listed by
\inlinecite{Nitta2013a} for 2010--2012, our estimations are close to
their results. Among these events, the 2012-09-20 flare turned out
to be the strongest one in the current solar cycle. Its probable
importance of about X13 was estimated from $L/R_\mathrm{S} \approx
2.38$. We also briefly discussed two far-side events of interest for
space weather. The 2012-07-23 eruption occurred in an active region
with a large magnetic flux and under favorable location relative to
the Earth could be a source of an extremely strong and prompt
geomagnetic storm. Judging from the B-streak, the west
behind-the-limb flare of 2014-01-06 was sufficiently intense and
long-lasting to be connected with a source of the observed proton
event including GLE.

Using the two LDE events as examples, we have demonstrated that by
measuring the maximum B-streak length in a number of consecutive
images one can reconstruct a probable time history of a flare. It is
clear that these and other features found due to B-streaks require
further detailed study. Hopefully, the proposed simple method of
prompt estimations of the SXR class of the far-side STEREO flares
from artifact B-streaks would be useful for solar studies and
solar-terrestrial forecasting.

\begin{acks}
We are grateful to an anonymous reviewer for constructive
comments, which helped us to improve the manuscript. The authors
thank the GOES and STEREO teams for their open data used in our
study. This research was supported by the Russian Foundation of
Basic Research under grant 14-02-00367 and the Ministry of
education and science of Russian Federation under projects 8407
and 14.518.11.7047.

\end{acks}

\end{article}

\end{document}